\documentclass[
 superscriptaddress,
 amsmath,amssymb,
 aps,
 floatfix
]{revtex4-2}

\usepackage{amsmath}
\usepackage{amssymb}
\usepackage{amsthm}
\usepackage{bm}
\usepackage{graphicx}
\usepackage{hyperref}
\usepackage[capitalise]{cleveref}
\usepackage[a-1b]{pdfx}

\newtheorem{conjecture}{Conjecture}

\begin{document}

\title{Spatial Dynamics of Higher Order Rock-Paper-Scissors and Generalisations}

\author{Christopher Griffin}
\email{griffinch@psu.edu}
\affiliation{
	Applied Research Laboratory,
    The Pennsylvania State University,
    University Park, PA 16802
    }
    
\author{Li Feng}
\email{fengl@cafs.ac.cn}
\affiliation{Fisheries Engineering Institute, Chinese Academy of Fishery Sciences, Beijing 100141, China}
    
\author{Rongling Wu}
\email{ronglingwu@bimsa.cn}
\affiliation{ Beijing Institute of Mathematical Sciences and Applications, Beijing 101408, China}
\affiliation{Yau Mathematical Sciences Center, Tsinghua University, Beijing 100084, China}

\date{\today~-~Preprint}

\begin{abstract} We introduce and study the spatial replicator equation with higher order interactions and both infinite (spatially homogeneous) populations and finite (spatially inhomogeneous) populations. We show that in the special case of three strategies (rock-paper-scissors) higher order interaction terms allow travelling waves to emerge in non-declining finite populations. We show that these travelling waves arise from diffusion stabilisation of an unstable interior equilibrium point that is present in the aspatial dynamics. Based on these observations and prior results, we offer two conjectures whose proofs would fully generalise our results to all odd cyclic games, both with and without higher order interactions, assuming a spatial replicator dynamic. Intriguingly, these generalisations for $N \geq 5$ strategies seem to require declining populations, as we show in our discussion.
\end{abstract}

\maketitle

\section{Introduction}
Replicator dynamics have been used extensively in theoretical ecology to model ecosystem interactions at a high level \cite{AL11,GBMA17,MCLM23}. Surprisingly, these models intersect those from theoretical physics, with tournament dynamics in ecology \cite{PG23,I87} also occurring in the analysis of Schr\"{o}dinger operator \cite{VS93} and in the discrete KdV equation \cite{B88}. 

Most biological and ecological models assume pairwise interactions \cite{M72,P81,RY02,KDGL22}, leading naturally to generalized Lotka-Volterra equations or the (equivalent) replicator dynamics in which the interaction matrix and the payoff matrix become synonymous. In this case, the payoff from interactions is used to define species fitness, as discussed in \cref{sec:Background}. This simple assumption is invalidated by strong evidence for the existence of higher order interactions  \cite{LBAA17,GBMA17,BKK16, MFFS19,SATA21,KKPL22,GLL22,BABB21,BCIL20,LRS19,SFF22}. Higher order interactions occur when three or more species (not necessarily distinct) interact with each other simultaneously to produce an additional payoff, which may increase or decrease fitness in the constituent species \cite{P81,LBAA17,GBMA17,BKK16, MFFS19,SATA21,MS17, MK19,DTS22}. In particular, higher order interactions have the potential  to alter the established relationship between diversity and stability \cite{BKK16}. 

While the replicator equation has been studied extensively \cite{W97,HS98,HS03,FS16}, the replicator dynamic with higher order interactions has recently been considered by Griffin and Wu \cite{GW23}. In this work, they show that the presence of higher order interactions in rock-paper-scissors can change the well-known dynamics of this game to allow the emergence of a sub-critical Hopf bifurcation as compared to the known degenerate Hopf bifurcation that characterizes the dynamics of rock-paper-scissors under the ordinary (pairwise) replicator dynamics \cite{Z80}. Before this, Gokhale and Traulsen \cite{GT10} studied evolutionary games with multiple (more than two) strategies and multiple players, while Zhang et al. \cite{ZPZW22} study multiplayer evolutionary games with asymmetric payoffs. In related but distinct work, Peixe and Rodrigues \cite{PR22a} study strange attractors and super-critical Hopf bifurcations in polymatrix replicators. Polymatrix games are also discussed in \cite{AD15,PG16}. However, to our knowledge, no one has yet studied a spatial replicator with higher order interactions, which is the goal of this paper.

Spatial evolutionary dynamics using partial differential equations have been studied by several authors, with \cite{CV97,dB13,KRFB02,NM92,RCS09,SSI04,V89,V91, HMT10,SMR14,PR17,PR19} providing a small example of the body of work. Most of these models assume an infinite, spatially homogeneous, population in so far as the state variables of the model are the proportions of the population playing a given strategy at a given location and time. Durrett and Levin were the first to point out the fundamental differences between discrete and continuous evolutionary game models and finite and infinite population assumptions \cite{DL94}. These distinctions have been further by Griffin et al. \cite{GMD21,G23}, where it is shown that finite populations can destroy travelling wave solutions (in rock-paper-scissors) or even reverse the direction of travelling waves (in prisoner's dilemma). 

Alternative approaches to studying finite populations frequently use discrete (grid) based methods and are based on the early work of Nowak and Martin \cite{NM92}, with extensions by several authors \cite{F08,KD96,M04,NBM94,ON06,P06,SH16,WH11,M94}. These models often focus on the interplay between concepts from statistical mechanics and evolutionary games via updating rules that use (among other mechanisms) the Boltzmann distribution. We will not consider these models in this paper.

Instead, we will use the models of Vickers \cite{V89} for infinite (or spatially homogeneous) populations and Griffin, Mummah and Deforest for finite (or spatially inhomogeneous) populations. In this paper, we study spatial replicator equations with higher order interactions for both infinite (spatially homogeneous) and finite (spatially inhomogeneous) populations. Formal definitions for these cases are provided in \cref{sec:Background}. While Griffin et al. \cite{GMD21} show that rock-paper-scissors under the ordinary spatial replicator dynamic can only admit travelling waves if the net population is decreasing, we show that the introduction of higher order interactions allows travelling waves to emerge in spatially homogeneous and inhomogeneous populations with no decline. Interestingly, when we generalise to cyclic games with more strategies (e.g., rock-paper-scissors-Spock-lizard), we see that this property of both the existence of travelling waves and a non-declining population seems to be a property of the three strategy case only. Nevertheless, we use observations made in this paper to pose two general conjectures on travelling waves and cyclic games under both the ordinary spatial replicator and the spatial replicator with higher order dynamics.

The remainder of this paper is organized as follows. In \cref{sec:Background}, we introduce notation needed in the remainder of the paper. We formulate the higher order spatial replicator in \cref{sec:HOInteractions}. Our analysis on rock-paper-scissors is carried out in \cref{sec:TravellingWave}. We generalise this analysis in \cref{sec:Generalisation}, proposing two conjectures on odd cyclic games. Conclusions and future directions are discussed in \cref{sec:Conclusion}. There is also an appendix (\ref{sec:FirstLyap}) that contains a derivation of the first Lyapunov coefficient for the Hopf bifurcation identified in the \cref{sec:TravellingWave}.

\section{Background}\label{sec:Background}
Let $\Delta_{n-1}$ be the $n-1$ dimensional unit simplex embedded in $\mathbb{R}^n$ composed of vectors $\mathbf{u} = \langle{u_1,\dots,u_n}\rangle$ where $u_1 + \cdots + u_n = 1$ and $u_i \geq 0$ for all $i=1,\dots,n$. We assume that an ecosystem supports a population size of $M$ total species. Then $u_iM$  is the size of the population of species $i$, where we allow fractional species counts for simplicity. 

Suppose the fitness of species $i$ is given by the function $f_i(\mathbf{u})$. The replicator equation with fitness $f$ is then,
\begin{equation*}
    \dot{u}_i = u_i\left[f_i(\mathbf{u}) - \bar{f}(\mathbf{u})\right],
\end{equation*}
where,
\begin{equation*}
    \bar{f}(\mathbf{u}) = \sum_j u_j f_j(\mathbf{u}),
\end{equation*}
is the mean fitness of the population. Assuming a finite population, the dynamics of the whole population are given by,
\begin{equation*}
    \dot{M} = \bar{f}(\mathbf{u})M.
\end{equation*}

If $\mathbf{A} \in \mathbb{R}^{n \times n}$ is a payoff (or interaction) matrix, then,
\begin{equation*}
    f_i(\mathbf{u}) = \mathbf{e}_i^T\mathbf{A}\mathbf{u}
\end{equation*}
produces the classic replicator from evolutionary game theory,
\begin{equation}
\dot{u}_i = u_i\left[\mathbf{e}_i^T\mathbf{A}\mathbf{u} - \mathbf{u}^T\mathbf{A}\mathbf{u}\right].
\label{eqn:AspatialReplicator}
\end{equation}

When $u$ is a function of space $\mathbf{x}$ and time $t$. Vickers \cite{V89,V91,HV92,VHB93,HV95,BPN11,NPB12} (and many others) study the spatial replicator with form,
\begin{equation}
\dot{u}_i = u_i\left[f_i(\mathbf{u}) - \bar{f}(\mathbf{u})\right] + D\nabla^2 u_i,
\label{eqn:Vickers}
\end{equation}
where $D$ is a diffusion constant. Without loss of generality, we assume that all species share a diffusion constant. Griffin, Mummah and DeForest \cite{GMD21} generalised the work of Durrett and Levin \cite{DL94} to show that when the total population $M(\mathbf{x},t)$ is neither homogeneous nor infinite, the species and total population are governed by the system of equations,
\begin{equation}
\left\{
\begin{aligned}
\frac{\partial u_i}{\partial t} =& u_i\left[f_i(\mathbf{u}) - \bar{f}(\mathbf{u}) 
    \right] + \frac{2nD}{M}\nabla M \cdot \nabla u_i + D\nabla^2 u_i.\\
\frac{\partial M}{\partial t} =& \bar{f}(\mathbf{u})M + \nabla^2 M,
\end{aligned}
\right.
\label{eqn:FiniteRep}
\end{equation}
where $n$ is the spatial dimension. It is straightforward to see that when $M(\mathbf{x},t)$ is homogeneous (or infinite), then the \cref{eqn:HOVickers} is recovered. As we will discuss these two cases in the remainder of the paper, we will refer to equations of the form given in \cref{eqn:FiniteRep} as the finite population spatial replicator and equations of the form given in \cref{eqn:Vickers} as the infinite population spatial replicator, even though we may really be considering finite populations that are spatially inhomogeneous vs. homogeneous. 

A biased rock-paper-scissors (RPS) payoff matrix is given by, 
\begin{equation}
\mathbf{A} = \left(\begin{array}{ccc}
    0 & -1 & 1+a\\
    1+a & 0 & -1\\
    -1 & 1+a & 0
\end{array} \right), 
\label{eqn:BiasedRPS}
\end{equation}
where we assume that $a > -2$ to maintain a rock-paper-scissors dynamics. It is well known that with this payoff matrix, the aspatial replicator, \cref{eqn:AspatialReplicator}, has a single interior fixed point at $u_1 = u_2 = u_3 = \tfrac{1}{3}$ and this fixed point is stable when $a > 0$, unstable when $a < 0$ and elliptic when $a = 0$ \cite{W97}.

Griffin, Mummah and DeForest \cite{GMD21} showed that a travelling wave solution exists for \cref{eqn:Vickers} using the biased RPS matrix just in case $a < 0$. In a finite population case, this is biologically unrealistic since,
\begin{equation*}
    \bar{f}(\mathbf{u}) = \mathbf{u}^T\mathbf{A}\mathbf{u} = a(u_1u_2 + u_1u_3 + u_2u_3),
\end{equation*}
which is negative just in case $a < 0$. Since, $\dot{M} = \bar{f}M < 0$ the population will collapse in this case. Moreover, \cite{GMD21} shows numerically that the travelling waves can be destroyed in the finite declining population case. However, we know that spatial travelling waves exist in real, non-declining populations \cite{BPCG14,K02,K92,SS08}. Our objective is to show that higher order interactions lead to the existence of travelling waves in cyclic competition (rock-paper-scissors) under both the finite and non-finite spatial replicator dynamics. 

\section{Higher Order Interactions in Rock-Paper-Scissors in Space} \label{sec:HOInteractions}
In \cite{GW23}, Griffin and Wu introduce a higher order interaction dynamic modelled by,
\begin{equation}
f_i(\mathbf{u}) = \mathbf{e}_i^T\mathbf{A}\mathbf{u} + \mathbf{u}^T\mathbf{B}_i\mathbf{u},
\label{eqn:f-HigherOrder}
\end{equation}
where $\mathbf{B}_i$ is a quadratic form (matrix) that takes two copies of the population proportion vector $\mathbf{u}$ and returns a payoff to species $i$ that occurs when one member of species $i$ randomly meets two members of the population. We think of $\mathbf{B}_i$ as being a slice of a $(0,3)$ tensor $\mathbf{B}:\Delta_{n-1}\times\Delta_{n-1}\times\Delta_{n-1} \to \mathbb{R}$. The mean fitness is then given by,
\begin{equation}
\bar{f} = \sum_{i = 1}^n u_i f_i(\mathbf{u}) = \sum_{i=1}^n u_i \left( \mathbf{e}_i^T\mathbf{A}\mathbf{u} + \mathbf{u}^T\mathbf{B}_i\mathbf{u} \right) = \\
\mathbf{u}^T\mathbf{A}\mathbf{u} + \sum_{i=1}^n u_i\mathbf{u}^T\mathbf{B}_i\mathbf{u}.
\label{eqn:fbar}
\end{equation}

Following Vickers, \cite{V91} we can construct a spatial model for higher order interactions that assumes a homogeneous population by appending a diffusion term to the replicator to obtain,
\begin{equation}
\frac{\partial u_i}{\partial t} = u_i\left( 
    \mathbf{e}_i^T\mathbf{A}\mathbf{u} - \mathbf{u}^T\mathbf{A}\mathbf{u} + \mathbf{u}^T\mathbf{B}_i\mathbf{u}  - \sum_{i=1}^n u_i\mathbf{u}^T\mathbf{B}_i\mathbf{u}
    \right) + D\nabla^2 u_i.
\end{equation}
Let $\mathbf{A}$ be the standard rock-paper-scissors matrix,
\begin{equation*}
\mathbf{A} = \left(\begin{array}{ccc}
    0 & -1 & 1\\
    1 & 0 & -1\\
    -1 & 1 & 0
\end{array} \right),
\end{equation*}
obtained by setting $a = 0$ in \cref{eqn:BiasedRPS}.
Now, $\mathbf{u}^T\mathbf{A}\mathbf{u} = 0$. Generalising from Griffin and Wu \cite{GW23}, we assume that the quadratic forms $\mathbf{B}_i$ ($i=1,2,3$) can be written as,
\begin{equation*}
    \mathbf{B}_1 = 
    \left(\begin{array}{ccc}
    0 & -\alpha & \beta\\
    -\alpha & -\gamma & 0\\
    \beta & 0 & \delta
\end{array} \right) \quad
\mathbf{B}_2 = 
    \left(\begin{array}{ccc}
    \delta & \beta & 0\\
    \beta & 0 & -\alpha\\
    0 & -\alpha & \gamma
\end{array} \right) \quad
\mathbf{B}_3 = 
    \left(\begin{array}{ccc}
    -\gamma & 0 & -\alpha\\
    0 & \delta & \beta \\
    -\alpha & \beta & 0
\end{array} \right),
\end{equation*}
where we assume, $\alpha,\beta,\gamma,\delta > 0$. As in \cite{GW23}, the tensor $\mathbf{B}$, composed of slices $\mathbf{B}_1$, $\mathbf{B}_2$ and $\mathbf{B}_3$, has cyclic structure. When we assume that $\gamma = 2\beta$ and $\delta = 2\alpha$, then,
\begin{equation*}
\sum_{i}u_i\mathbf{u}^T\mathbf{B}_i\mathbf{u} = 0,
\end{equation*}
and consequently, $\bar{f} = 0$. Thus, any finite population would be stable assuming these dynamics. The resulting spatial dynamics for a homogeneous (infinite) population are,
\begin{equation}
\frac{\partial u_i}{\partial t} = u_i\left( 
    \mathbf{e}_i^T\mathbf{A}\mathbf{u} + \mathbf{u}^T\mathbf{B}_i\mathbf{u} 
    \right) + D\nabla^2 u_i.
\label{eqn:HOVickers}
\end{equation}
The corresponding finite population model is then,
\begin{equation}
\left\{
\begin{aligned}
\frac{\partial u_i}{\partial t} =& 
u_i\left( \mathbf{e}_i^T\mathbf{A}\mathbf{u} + \mathbf{u}^T\mathbf{B}_i\mathbf{u} \right) + \frac{2nD}{M}\nabla M \cdot \nabla u_i + D\nabla^2 u_i\\
\frac{\partial M}{\partial t} =& D \nabla^2 M.
\end{aligned}
\right.
\label{eqn:HOFiniteRep}
\end{equation}
Notice that our assumption on $\mathbf{A}$ and $\mathbf{B}$ implies that $\bar{f} = 0$ and so the bulk population is governed by the diffusion equation. 

\section{Travelling Wave Solutions exist in One Dimension}\label{sec:TravellingWave}
We begin by analysing the aspatial dynamics. Under the assumption that $\gamma = 2\beta$ and $\delta = 2\alpha$, the aspatial dynamics are given by,
\begin{align}
\dot{u}_1 & = u_1
   \left(u_3 - u_2\right) + 2 u_1 \left(\alpha  u_3^2 +\beta  u_1 u_3 -\alpha  u_2 u_1-\beta  u_2^2\right)\label{eqn:ur}\\
\dot{u}_2 & = u_2\left(u_1-u_3\right) + 2 u_2 \left(\alpha  u_1^2 + \beta  u_2 u_1-\alpha  u_2 u_3-\beta  u_3^2\right)\label{eqn:up}\\
\dot{u}_3 & = u_3\left(u_2-u_1\right) + 2 u_3 \left(\alpha  u_2^2 + \beta  u_2 u_3-\alpha  u_1 u_3-\beta  u_1^2\right)\label{eqn:us}
\end{align}
Just as with ordinary rock paper scissors, the three extreme points of $\Delta_2$ are fixed points as is the interior point $u_1 = u_2 = u_3 = \tfrac{1}{3}$. First order analysis of the Jacobian matrix at the interior fixed point gives eigenvalues,
\begin{align}
    \lambda_1 &= 0\\
    \lambda_{2,3} &= \frac{\beta-\alpha}{3} \pm i\frac{\sqrt{3}}{3}(1+\alpha + \beta).
\end{align}
Thus the interior fixed point is unstable when $\beta > \alpha$ and stable if $\beta < \alpha$. When $\beta = \alpha$, the Hartman-Grobman theorem cannot be used. In this case, the dynamics simplify to,
\begin{align*}
\dot{u}_1 & = (1+2\alpha) u_1 \left(u_3 - u_2\right) \\
\dot{u}_2 & = (1+2\alpha)u_2\left(u_1-u_3\right) \\
\dot{u}_3 & = (1+2\alpha)u_3\left(u_2-u_1\right),
\end{align*}
which is just an ordinary rock-paper-scissors dynamic with payoff matrix $(1+2\alpha)\mathbf{A}$. Therefore, the interior fixed point is elliptic in this case. Moreover, we have shown that the higher order dynamics we consider have analogous dynamics to the ordinary RPS system, except that by construction $\bar{f} = 0$. 

Now consider the spatial replicator with infinite (homogeneous) population. Let $z = x + ct$, where $c$ is a wave speed to be determined. If we have $\mathbf{u}(x,t) = \mathbf{u}(z)$, then the resulting system becomes,
\begin{equation}
c u'_i = u_i\left(\mathbf{e}_i^T\mathbf{A}\mathbf{u} + \mathbf{u}^T\mathbf{B}_i\mathbf{u}\right) + D u_i'',
\end{equation}
where $u'_i$ is the derivative in terms of $z$. Let $v_i = u'_i$. Then we have the modified system of differential equations,
\begin{equation}
\left\{
\begin{aligned}
D v_i' &= cv_i - u_i\left(\mathbf{e}_i^T\mathbf{A}\mathbf{u} + \mathbf{u}^T\mathbf{B}_i\mathbf{u}\right)\\
u_i' &= v_i.
\end{aligned}
\right.
\label{eqn:TravellingWave}
\end{equation}
This system has a fixed point at $v_i = 0$, $u_i = \tfrac{1}{3}$ for $i=1,2,3$. Computing the eigenvalues of the Jacobian at this point gives,
\begin{align*}
\lambda_1 &= 0\\
\lambda_2 &= \frac{c}{D}\\
\lambda_{3,4} &= \frac{3c \pm \sqrt{9c^2-12D(\beta - \alpha) + 12Di\sqrt{3}(1+\alpha + \beta)}}{6D}\\
\lambda_{5,6} &= \frac{3c \pm \sqrt{9c^2-12D(\beta - \alpha) - 12Di\sqrt{3}(1+\alpha + \beta)}}{6D}.
\end{align*}
The zero eigenvalue arises because we necessarily have $u_r(z) + u_p(z) + u_s(z) = 1$ and $v_r(z) + v_p(z) + v_s(z) = 0$ and thus the dynamics play out on a $4$ dimensional manifold and $\lambda_1$ and $\lambda_2$ can be ignored.

Focusing on the term under the outer radical, assume there is some $r$ so that, 
\begin{equation*}
(3c\pm r i)^2 = 9c^2-12D(\beta - \alpha) \pm 12Di\sqrt{3}(1+\alpha + \beta).
\end{equation*}
Then we obtain the equations,
\begin{align*}
&9c^2 - r^2 = 9c^2 - 12D(\beta - \alpha)\\
&6cr = 12D\sqrt{3}(1+\alpha+\beta).
\end{align*}
We can compute the wave speed and the parameter $r$ as,
\begin{equation*}
(r^\star, c^\star) = \pm \left( \sqrt{12D(\beta - \alpha)},\frac{D(1+\alpha + \beta)}{\sqrt{D(\beta - \alpha)}}\right).
\end{equation*}
We conclude that the wave speed is non-imaginary, just in case $\beta > \alpha$. That is, a travelling wave can emerge when the interior fixed point of the aspatial dynamics is unstable and hence stabilised by the diffusion term. This is similar to the condition found by Griffin, Mummah and Deforest for the ordinary spatial replicator with rock-paper-scissors \cite{GMD21}. 

We can simplify the eigenvalues $\lambda_{3,4}$ and $\lambda_{5,6}$ using the negative branch of $(r^\star,c^\star)$ to obtain,
\begin{align*}
\lambda_{4,6} =& \pm\frac{i}{\sqrt{3}}\sqrt{\frac{\beta -\alpha }{D}}\\
\lambda_{3,5} = & -\frac{(1+\alpha+\beta)}{\sqrt{D (\beta -\alpha )}} \pm i\frac{\sqrt{\beta - \alpha}}{\sqrt{3D}}.
\end{align*}
Thus we have three eigenvalues with negative real part indicating a three-dimensional stable manifold with two additional eigenvalues that are pure imaginary. The presence of a stable manifold with imaginary eigenvalues satisfies the first criterion of Hopf's theorem \cite{GH13} (Page 152). We use the negative branch because that will ensure that solutions to the PDE are (locally) attracted to the limit cycle and hence the travelling wave solution.

Now consider the specific eigenvalues,
\begin{equation*}
    \lambda_{4,6} = \frac{3c - \sqrt{9c^2-12D(\beta - \alpha) \pm 12Di\sqrt{3}(1+\alpha + \beta)}}{6D}.
\end{equation*}
Differentiating with respect to $c$ and evaluating at the identified wave speed yields,
\begin{equation*}
     \lambda_{4,6}'(c^\star) = \frac{1}{2D} - \frac{c^\star\sqrt{3}}{2D\sqrt{(3c^\star \pm {r^\star}i)^2}} = \frac{1}{2D} - \frac{\sqrt{3}c^\star(3c^\star\mp {r^\star} i)}{2D(9{c^\star}^2+{r^\star}^2)}.
\end{equation*}
Then,
\begin{equation*}
    Re[\lambda'_{4,6}(c^\star)] = \frac{1}{2D}\left(
    1-\frac{\sqrt{3} (\alpha +\beta +1)^2}{7 \alpha ^2-2 \alpha  (\beta -3)+\beta  (7 \beta +6)+3}
    \right)
\end{equation*}
To see that this is always non-zero, note that the equation,  
\begin{equation*}
    1-\frac{\sqrt{3} (\alpha +\beta +1)^2}{7 \alpha ^2-2 \alpha  (\beta -3)+\beta  (7 \beta +6)+3} = 0,
\end{equation*}
is quadratic in $\alpha$ and $\beta$. Solving for $\alpha$ in terms of $\beta$ leads to a quadratic equation with discriminant,
\begin{equation*}
    \mathcal{D} = 16 \left(\sqrt{3}-3\right) (2 \beta +1)^2 < 0. 
\end{equation*}
Thus, there are no real values of $\alpha$ and $\beta$ that make this expression zero. As such, the eigenvalues must cross the imaginary axis with non-zero speed, satisfying the second criterion of Hopf's theorem. Thus, we have proved the existence of a Hopf bifurcation at the fixed point, which implies the existence of an isolated attracting period orbit (stable limit cycle) just in case the first Lyapnuov coefficient of the system's normal form is non-zero and negative \cite{GH13}. The first Lyapunov coefficient can be constructed using techniques in \cite{KKK98,PCL12}, as,
\begin{equation*}
    \ell_1 = -\frac{3 \sqrt{3} \left(6 D \xi  \omega ^2+\xi ^2 -3 D^2 \omega ^4\right)}{4 \sqrt{\left(\omega ^2+1\right)^3 \left(3 D^2 \omega ^4+\xi ^2\right) \left(3 D^2 \omega ^2
   \left(\omega ^2+1\right)+\xi ^2\right)}},
\end{equation*}
where,
\begin{equation*}
\omega = \sqrt{\frac{\beta-\alpha}{3D}} \quad \text{and} \quad
\xi = 1 + \alpha + \beta.
\end{equation*}

The details of this construction are provided in \ref{sec:FirstLyap} and the SI, where it is also shown that this quantity is always negative. Thus, by Hopf's theorem, we have proved that the travelling wave system \cref{eqn:TravellingWave} has an attracting periodic solution (because $\ell_1 < 0$) and consequently a travelling wave solution must exist for \cref{eqn:HOVickers}.

We now consider the one-dimensional finite population model from \cref{eqn:HOFiniteRep}. In one dimension we have,
\begin{equation*}
\left\{
\begin{aligned}
\partial_t u_1 & = u_1
   \left(u_3 - u_2\right) + 2 u_1 \left(\alpha  u_3^2 +\beta  u_1 u_3 -\alpha  u_2 u_1-\beta  u_2^2\right) + \frac{2D}{M}\partial_x M\partial_x u_1  + D\partial_{xx} u_1\\
\partial_t u_2 & = u_2\left(u_1-u_3\right) + 2 u_2 \left(\alpha  u_1^2 + \beta  u_2 u_1-\alpha  u_2 u_3-\beta  u_3^2\right) + \frac{2D}{M}\partial_x M\partial_x u_2  + D\partial_{xx} u_2\\
\partial_t u_3 & = u_3\left(u_2-u_1\right) + 2 u_3 \left(\alpha  u_2^2 + \beta  u_2 u_3-\alpha  u_1 u_3-\beta  u_1^2\right) + \frac{2D}{M}\partial_x M\partial_x u_3  + D\partial_{xx} u_3\\
\partial_t M &= D \partial_{xx}M.
\end{aligned}
\right.
\end{equation*}
Following work by Griffin \cite{G23}, we have a travelling wave solution for the diffusion equation, $\partial_t M = D \partial_{xx}M$ as,
\begin{equation}
    M(x,t) = A\exp\left[c(x+ Dct)\right] + B,
\label{eqn:PopulationTravelingWave}
\end{equation}
where $A$ and $B$ are arbitrary constants and $kc \in \mathbb{R}$ is the population wave speed, and $c\in \mathbb{R}$ will be the species wave speed. Assume $B = 0$. Then
\begin{equation}
\frac{1}{M}\frac{\partial M}{\partial x} =  c.
\end{equation} 
Then in the finite dimensional case, the resulting travelling wave equation for the species is, 
\begin{equation*}
c u'_i = u_i\left(\mathbf{e}_i^T\mathbf{A}\mathbf{u} + \mathbf{u}^T\mathbf{B}_i\mathbf{u}\right) + 2Dc u'_i + D u_i'',
\end{equation*}
leading to the system of equations,
\begin{align*}
D v_i' &= c(1-2D)v_i - u_i\left(\mathbf{e}_i^T\mathbf{A}\mathbf{u} + \mathbf{u}^T\mathbf{B}_i\mathbf{u}\right)\\
u_i' &= v_i.
\end{align*}
This is identical to \cref{eqn:TravellingWave} but with a modified wave speed and consequently, our proof of the existence of a travelling wave solution applies \textit{mutatis mutandis}. Thus, for small diffusion ($D < \tfrac{1}{2})$, the finite and infinite populations will share solutions but travelling at different speeds. 

\begin{figure}[htbp]
\centering
\includegraphics[width=0.95\textwidth]{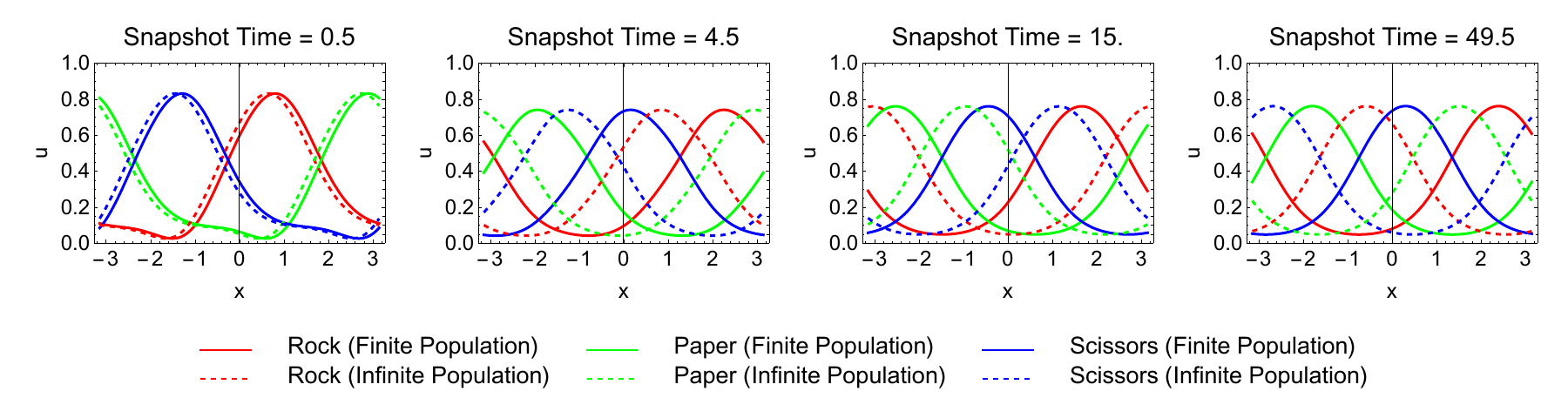}\\
\caption{Both the finite and infinite population spatial replicator with higher order interactions in RPS converge to a travelling wave solution with distinct wave speeds depending on whether a finite and spatially inhomogeneous or spatially homogeneous (or infinite) population is modelled. }
\label{fig:TravellingWave}
\end{figure}

We now illustrate this for $D = \tfrac{1}{10}$, $\beta = \tfrac{3}{2}$ and $\alpha = 1$, and simultaneously show the existence of the predicted limit cycle solution for \cref{eqn:TravellingWave}. Consider the following initial conditions for the PDE's \cref{eqn:HOVickers} and \cref{eqn:HOFiniteRep},
\begin{align*}
    u_1(x,0) &= \frac{1}{3}\left[1+\sin(x)\right],\\
    u_2(x,0) &= \frac{1}{3}\left[1 + \sin\left(x - \frac{2\pi}{3}\right)\right],\\ 
    u_3(x,0) &= \frac{1}{3}\left[1 + \sin\left(x - \frac{4\pi}{3}\right)\right],
\end{align*}
and assume periodic boundary conditions $u_i(-\pi,t) = u_i(\pi,t)$. Then four snapshots of the resulting travelling wave solution are shown in \cref{fig:TravellingWave}.
We see a perturbation of the initial condition that quickly settles into the travelling wave solution in both the finite and infinite population cases. 

\begin{figure}[htbp]
\centering
\includegraphics[width=0.45\textwidth]{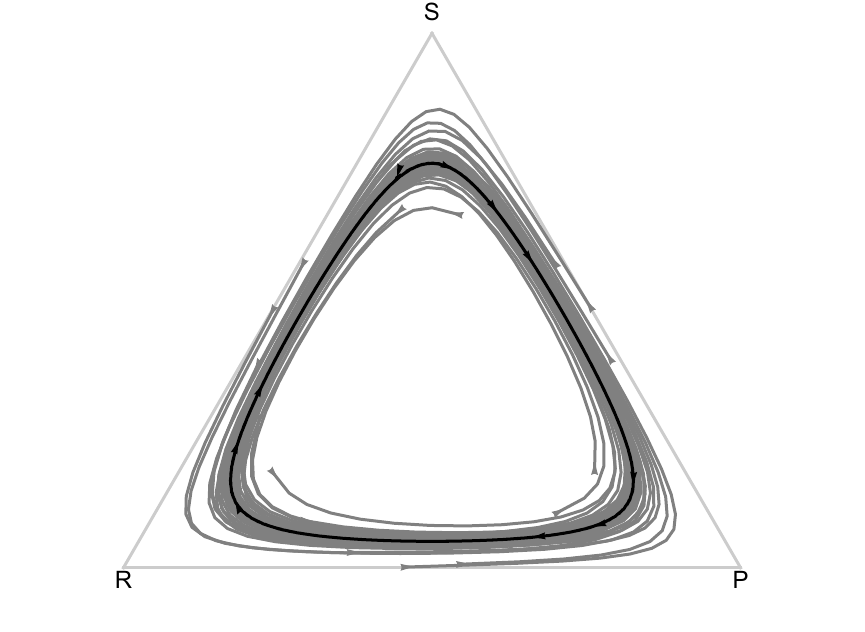} \quad
\includegraphics[width=0.45\textwidth]{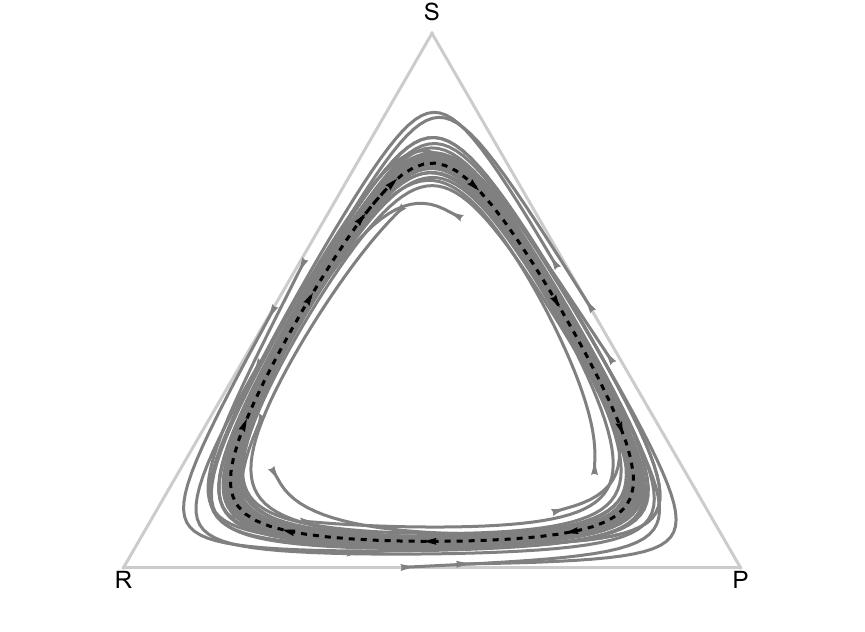}
\caption{The PDE solutions converge to a travelling wave solution, which corresponds to an attracting limit cycle in the equivalent system of ordinary differential equations. The (left) shows the spatially homogeneous or infinite population spatial replicator with higher order interactions in RPS. The (right) shows the finite spatially inhomogeneous population spatial replicator with higher order interactions in RPS.}
\label{fig:LimitCycle}
\end{figure}

We can numerically investigate solutions for \cref{eqn:TravellingWave}. Let $u_i(x,t)$ be the (numerical) travelling wave solutions to the infinite (finite) population spatial replicator. For $u_i(z)$ and $v_i(z)$ in \cref{eqn:TravellingWave}, we set,
\begin{align*}
    u_i(0) &= u_i(0,T)\\
    v_i(0) &= \partial_x u_i(0,T),
\end{align*}
where $T = 50$ is sufficiently large to ensure that the resulting numeric solution is (effectively) on the limit cycle. When we plot $u_i(0,t)$ for $i=1,2,3$ (in an appropriate projection) we see that the solutions to the partial differential equation(s) approach the limit cycle, as expected. This is shown in \cref{fig:LimitCycle} 

Recall that when $\beta > \alpha$, the interior fixed point $u_i = \tfrac{1}{3}$ ($i=1,2,3$) in the aspatial dynamics is unstable. We conclude that the travelling wave solution arises because the diffusion is stabilising the growing oscillations that would arise at all points in space and (under certain initial conditions), allowing the stabilised oscillations to synchronize. 

We can prove this stabilisation occurs by first order analysis of the infinite population system. Let $\mathbf{J}_0$ be the Jacobian of the equation system given in \crefrange{eqn:ur}{eqn:us} evaluated at the interior fixed point. Then we have,
\begin{equation}
    \mathbf{J}_0 = \begin{bmatrix}
        \frac{2}{9} (\beta - \alpha) & -\frac{1}{9} (2 \alpha + 4 \beta +3) & \frac{1}{9} (4
   \alpha +2 \beta +3) \\
 \frac{1}{9} (4 \alpha +2 \beta +3) & \frac{2}{9} (\beta - \alpha ) & -\frac{1}{9} (2 \alpha +4 \beta +3) \\
 -\frac{1}{9} (2 \alpha +4 \beta +3) & \frac{1}{9} (4 \alpha +2 \beta +3) & \frac{2}{9}
   (\beta - \alpha )
    \end{bmatrix}.
\end{equation}
Let $\upsilon_i = u_i - \tfrac{1}{3}$ with $\bm{\upsilon} = \langle{u_1,u_2,u_3}\rangle$ and let $\mathbf{D} = D\mathbf{I}$, where $\mathbf{I}$ is the identify matrix. Following \cite{MM03}, we analyse the linearised stationary problem with Neumann boundary conditions, 
\begin{equation*}
\mathbf{0} = \mathbf{J}_0\bm{\bm\upsilon} + \mathbf{D}\nabla^2\bm{\upsilon}, \quad (\mathbf{n} \cdot \nabla)\bm{\upsilon} = 0,
\end{equation*}
by computing the roots of the characteristic polynomial,
\begin{equation*}
    \mathrm{det}\left(\lambda\mathbf{I} + \mathbf{J}_0 + \mathbf{D}k^2\right).
\end{equation*}
Here $k$ is a wave number in a Fourier basis of a proposed solution ansatz and $\lambda$ is an eigenvalue. We find three eigenvalues,
\begin{align*}
    &\lambda_1 = -Dk^2\\
    &\lambda_{2,3} = \frac{1}{3}\left(\beta - \alpha -3Dk^2 \pm i\sqrt{3(1+\alpha + \beta)}\right).
\end{align*}
The fact that $-Dk^2$ appears in the real parts of all three eigenvalues is sufficient to show that the diffusion exerts only a stabilising effect. Moreover, 
\begin{equation*}
    \mathrm{Re}(\lambda_{2,3}) = \frac{\beta - \alpha - 3Dk^2}{3},
\end{equation*}
is positive only if $\beta > \alpha + 3Dk^2$. That is $\beta > \alpha$, which we already knew. Thus, we have not only shown that the diffusion exerts a stabilising effect on the system, but also that Turing patterns cannot emerge in this system as a result of diffusion induced instability. Interestingly, this seems also explain the occurrence of travelling waves when no higher order interactions are present but $a < 0$ in the interaction matrix in \cref{eqn:BiasedRPS} using the infinite population spatial replicator as shown in \cite{GMD21}. We discuss this as a possible framework for generalising these results in future directions.

\begin{figure}[htbp]
\centering
\includegraphics[width=0.9\textwidth]{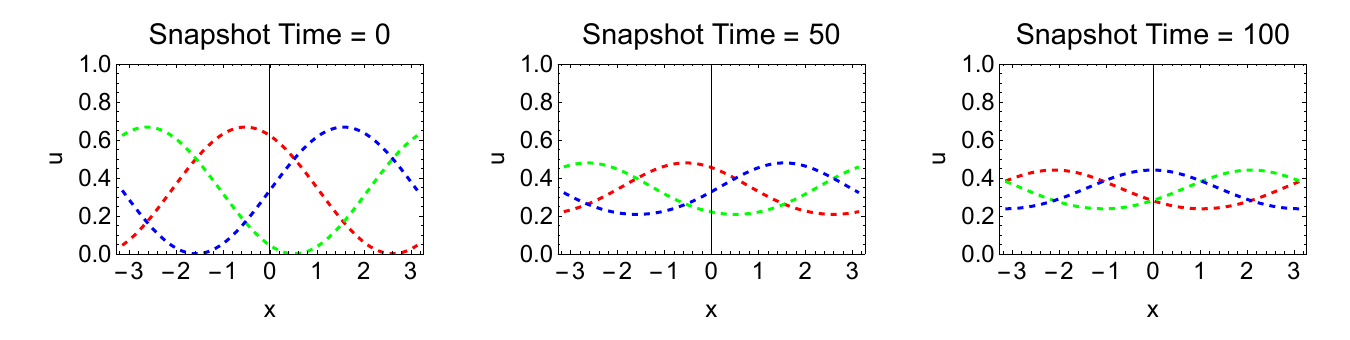}
\caption{The amplitude of the travelling wave solution decreases, approaching the fixed point solution $u_i(x,t) = \tfrac{1}{3}$ as $k$ increases.}
\label{fig:SmallerLimitCycle}
\end{figure}

While it is generally difficult to construct the amplitude of a limit cycle, and thus a travelling wave, we can show numerically that the amplitude of the travelling wave (limit cycle) varies inversely with (a function of) the diffusion constant. Thus, as $D$ increases, we expect to see travelling wave solutions that approach the fixed point solution $u_i(x,t) = \tfrac{1}{3}$, further demonstrating the stabilising effect of the diffusion. This is illustrated in \cref{fig:SmallerLimitCycle} for $D = \tfrac{1}{6} < \tfrac{1}{10}$ in the infinite population case.

\begin{figure}[htbp]
\centering
\includegraphics[width=0.95\textwidth]{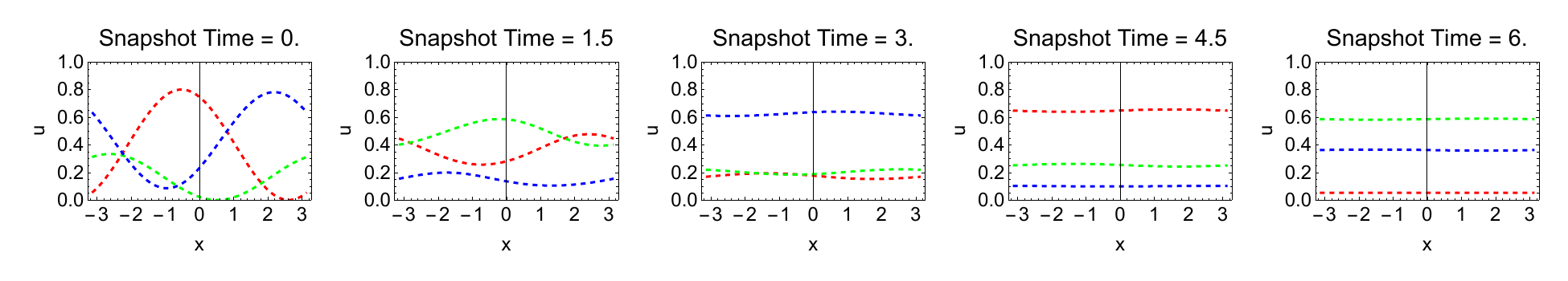}\\
\includegraphics[width=0.95\textwidth]{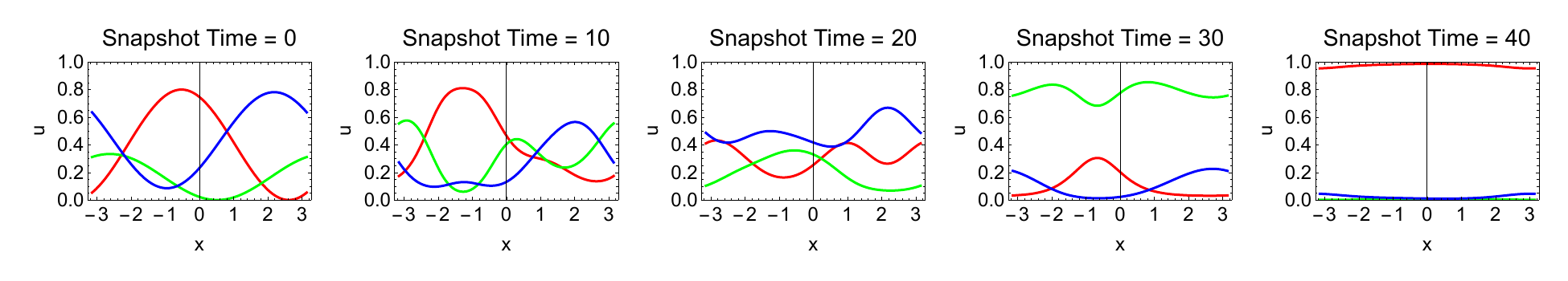}
\caption{(Top) The infinite population dynamics with initial conditions that do not lead to a travelling wave. (Bottom) The finite population dynamics with initial conditions that do not lead to a travelling wave. Notice the non-homogeneity of the population preserves spatially non-homogeneous species dynamics longer.}
\label{fig:NonWave}
\end{figure}

We can prove, by counter-example, that the travelling wave solution is not globally asymptotically stable in the space of solutions for either the finite population equation or the infinite population equation. To see this, consider the initial condition,
\begin{align*}
    u_1(x,0) &= \frac{2}{5} \left[\cos \left(x+\frac{\pi }{6}\right)+1\right],\\
    u_2(x,0) &= \frac{1}{6} \left[1-\cos \left(\frac{\pi }{6}-x\right)\right],\\ 
    u_3(x,0) &= \frac{1}{60} \left[17 \sin (x)-7 \sqrt{3} \cos (x)+26\right],
\end{align*}
These expressions do not lead to (numerical) solutions that tend to travelling waves, as shown in \cref{fig:NonWave}.
Instead, these solutions lead to a globally oscillating solution that asymptotically approaches the boundary of the simplex at all spatial positions. For the finite population case, we are using the travelling wave solution used in our prior numerical illustration. Interestingly, the finite population case takes longer to approach the globally oscillating solution than the infinite population case, most likely as a result of the bulk movement of the finite population. This is illustrated in \cref{fig:NonWave}(bottom). This phenomenon may warrant investigation in future work.

\section{Generalisation}\label{sec:Generalisation}
To generalise the work in this paper, recall that a circulant matrix \cite{D12} has structure,
\begin{displaymath}
\mathbf{A} =
\begin{bmatrix}
a_0 & a_{n-1} & a_{n-2} & \cdots & a_{1}\\
a_1 & a_0 & a_{n-1} & \cdots & a_{2}\\
\vdots & \vdots & \vdots & \ddots & \vdots\\
a_{n-1} & a_{n-2} & a_{n-3} & \cdots & a_0
\end{bmatrix}.
\end{displaymath}
That is, the entire matrix structure is determined by the first row. The set of all circulant matrices forms an algebra under addition and (commutative) matrix multiplication. 

Let $N = 2n+1$ with $n \geq 1$. Consider the $N$ dimensional row vector,
\begin{equation*}
    \mathbf{A}_{N_1} = \begin{bmatrix}
        0, -1, 1+a, -1, 1+a, \dots, -1, 1+a
    \end{bmatrix},
\end{equation*}
where $a$ is the biasing term. Then the $N\times N$ circulant matrix $\mathbf{A}_N$ defined by $\mathbf{A}_{N_1}$ is the payoff matrix for the $N$ strategy generalisation of rock-paper-scissors. Griffin and Fan \cite{GF22} showed that the replicator dynamics \cref{eqn:AspatialReplicator} have a unique interior fixed point at $\mathbf{u} = \left\langle{\tfrac{1}{N},\dots,\tfrac{1}{N}}\right\rangle$ that is stable when $a > 0$ and unstable when $a < 0$. Then we have the following conjecture, which is proved for the case $N = 3$.
\begin{conjecture} For all odd $N$, the one-dimensional spatial replicator equation,
\begin{equation*}
    \dot{u}_i = u_i\left(\mathbf{e}_i^T\mathbf{A}_N\mathbf{u} - \mathbf{u}\mathbf{A}_N\mathbf{u}\right) + D\frac{\partial u_i}{\partial x},
\end{equation*}
admits a travelling wave solution when $\mathbf{A}_N$ is defined as above and $a < 0$.
\end{conjecture}
We hypothesize that 
As in the three strategy case (rock-paper-scissors), when $a < 0$, $\mathbf{u}\mathbf{A}_N\mathbf{u} < 0$, which implies a globally decreasing population.

Generalising our result to the higher order interaction case produces a surprising result. To generalise the interaction tensor to the case of $N$ strategies, let $\Sigma = \{1,\dots,N\}$ be the strategy set and let $w(i)$ denote the set of strategies that are beaten by strategy $i$ and let $l(i)$ be the set of strategies that beat strategy $i$. Then,
\begin{equation}
    \mathbf{B}_{N_{i_{jk}}} = 
    \begin{cases}
        \delta & \text{if $j,k \in w(i)$}\\
        -\gamma & \text{if $j,k \in l(i)$}\\
        \beta & \text{if $j = i$ and $k \in w(i)$ or $k = i$ and $j \in w(i)$}\\
        -\alpha & \text{if $j = i$ and $k \in l(i)$ or $k = i$ and $j \in l(i)$}\\
        0 & \text{$j \in w(i)$ and $k \in w(j)$ and $i \in w(k)$}\\
        0 & \text{$j \in l(i)$ and $k \in l(j)$ and $i \in l(k)$}\\
        0 & \text{$i = j = k$}.
    \end{cases}
    \label{eqn:HOMatrix}
\end{equation}
The last three cases produce the $0$ diagonals and the case when the three strategies form a winning/losing cycle (like rock, paper and scissors). As before, $\alpha, \beta, \gamma, \delta > 0$. When $N = 3$, we recover the higher order interaction matrices we have already studied.

Consider $N = 5$. Then evaluating at $\gamma = 2\beta$ and $\delta = 2\alpha$ gives,
\begin{equation*}
    \bar{f} = \sum_{i} u_i\mathbf{u}\mathbf{B}_{N_i}\mathbf{u} = 4 (\alpha -\beta
   )\left(u_1 u_2 u_4+u_1 u_3 u_4+u_2 u_5 u_4+u_1 u_3 u_5+u_2 u_3 u_5\right). 
\end{equation*}
This value is $0$ if and only if $\alpha = \beta$ and otherwise its sign is equivalent to $\mathrm{sgn}(\alpha - \beta)$. Notice, the triples are composed of entries of the form $u_iu_ju_k$ where $j,k \in l(i)$ and therefore cannot occur in the case when $N = 3$, which is why $\bar{f} = 0$ in our prior analysis. Simple computation shows that the only way for $\bar{f}$ to be zero is in the case when $\alpha = \beta$. Further analysis shows that the eigenvalues of the Jacobian at the (unique) interior fixed point $\mathbf{u}_i = \tfrac{1}{5}$ ($i=1,\dots,5$) are,
\begin{align}
    &\lambda_1 = \frac{4}{25}(\beta - \alpha)\\
    &\lambda_{2,3} = \frac{1}{25} \left(7(\beta-\alpha) \pm 5 i (1 + \alpha +\beta)\sqrt{5+2 \sqrt{5}}  \right)\\
    &\lambda_{4,5} = \frac{1}{25} \left(7(\beta-\alpha) \pm 5 i (1 + \alpha +\beta)\sqrt{5-2 \sqrt{5}}  \right).
\end{align}
Thus the interior fixed point is unstable just in case $\beta < \alpha$, as in the three strategy case, which we conjecture will lead to a stable travelling wave solution in the infinite population spatial case. We summarize this in the following conjecture.
\begin{conjecture} For all odd $N$, the one-dimensional higher order spatial replicator equation,
\begin{equation*}
\frac{\partial u_i}{\partial t} = u_i\left( 
    \mathbf{e}_i^T\mathbf{A}_N\mathbf{u} - \mathbf{u}^T\mathbf{A}_N\mathbf{u} + \mathbf{u}^T\mathbf{B}_{N_i}\mathbf{u}  - \sum_{i=1}^n u_i\mathbf{u}^T\mathbf{B}_{N_i}\mathbf{u}
    \right) + D\nabla^2 u_i,
\end{equation*}
admits a travelling wave solution when $\mathbf{A}_N$ and $\mathbf{B}_{N_i}$ ($i=1,\dots,N$) are defined as above and $a = 0$, $\gamma = 2\beta$ and $\delta = 2\alpha$ and $\beta > \alpha$.  
\end{conjecture}
We note that the computation of the first Lyapunov coefficient will most likely be the most difficult component of any proofs of the generalised conjectures. 

What is perhaps the most interesting aspect of this is the fact that higher order interactions as defined by \cref{eqn:HOMatrix} seem to be able to simultaneously produce travelling wave solutions and maintain a constant population size only for the three strategy rock-paper-scissors game. While we do not rule out the possibility that a more complex interaction mechanism may be able to simultaneously accomplish this, it is surprising that this property seems to hold for only the three strategy case and thus may warrant additional study. 

\section{Conclusions and Future Directions}\label{sec:Conclusion}
In this paper, we merged the higher order interaction model first discussed by Griffin and Wu \cite{GW23} with the spatial replicator equation model of Vickers \cite{V91} and the finite population spatial model of Griffin, Mummah and Deforest \cite{GMD21}. For higher order interactions in rock-paper-scissors, we showed that travelling wave solutions exist in both the finite and infinite population cases, with the important model feature that the net population was stable (as opposed to declining). This suggests that if replicators are models of real-world cyclically interacting populations, then travelling waves in such populations can be explained by either a declining population count or the presence of higher order (i.e., non-pairwise) interactions. In discussing a generalisation of this approach, we provided two conjectures on the existence of travelling wave solutions in spatial replicator dynamics with an arbitrary odd number of strategies. Most interestingly, we found that the property of population size preservation and the existence of travelling wave solutions appears to be present only in the rock-paper-scissors game (three strategy case), with higher order interactions. Games with more than three strategies (e.g., rock-paper-scissors-Spock-lizard) seem to admit travelling wave solutions only when the total population is decreasing and higher order interactions cannot remediate this.

Proving the conjectures raised in this paper is clearly an area of future work. We argue that stable Turing patterns will not be admitted by the infinite population spatial replicator with higher order interactions as defined in this paper. However, Griffin and Wu \cite{GW23} show that a (subcritical) Hopf bifurcation can emerge in the aspatial higher order dynamics using a related but distinct payoff matrix and higher order interaction matrices. If parameter regimes exist where a supercritical Hopf bifurcation exists in the aspatial case, then it may be possible that a diffusion mediated transition maybe possible from periodic solutions to asymptotically stable solutions as in the work of Ginzburg and Landau equation \cite{GLSS13} or in the work of  Dil\~{a}o \cite{D05}. Investigating this possibility of significant interest for future work. 

\section*{Acknowledgements}
C.G. was supported in part by the National Science Foundation under grant CMMI-1932991. C.G. would also like to thank Andrew Belmonte for a useful discussion on this topic.

\section*{Data and Code Availability}
Three Mathematica notebooks are provided as supplementary materials and contain the code needed to reproduce the images and theoretical derivations in this paper.

\appendix
\section{Constructing the First Lyapunov Coefficient}\label{sec:FirstLyap}
The approach outlined here is provided in  \cite{PCL12} and is distilled from the detailed discussion in \cite{KKK98}. We begin by setting $u_s = 1 - u_r - u_p$ and $v_s = -v_r - v_p$, since we can see that $v_r + v_p + v_s = 0$. Then \cref{eqn:TravellingWave} reduces to four linearly independent equations. Let $\mathbf{J}_0$ be the Jacobian of this reduced dimension system evaluated at the fixed point $u_r = u_p = \tfrac{1}{3}$ and $v_r = v_p = 0$ and the special wave speed $c = c^\star$ using the negative branch. Thus,
\begin{equation*}
    \mathbf{J}_0 = \begin{bmatrix}
    \frac{\xi}{D\omega\sqrt{3}} & \frac{\xi-3D\omega^2}{3 D} & 0 & \frac{2 \xi}{3 D} \\
 1 & 0 & 0 & 0 \\
 0 & -\frac{2 \xi}{3 D} & \frac{\xi}{D\omega\sqrt{3}} & -\frac{3D\omega^2+\xi}{3 D} \\
 0 & 0 & 1 & 0 
    \end{bmatrix}
\end{equation*}
As expected, this matrix  has two pure imaginary eigenvalues of form $\pm\omega i$, where,
\begin{equation*}
    \omega = \sqrt{\frac{\beta - \alpha}{3D}} \quad \text{and} \quad \xi = 1 + \alpha + \beta.
\end{equation*}
Let $\mathbf{q}$ be the normalized eigenvector of $\mathbf{J}_0$ so that $\mathbf{J}_0\mathbf{q} = i\omega\mathbf{q}$ and let $\mathbf{p}$ be the normalized eigenvector of $\mathbf{J}_0^T$ so that $\mathbf{J}_0^T\mathbf{p} = -i\omega\mathbf{p}$. The values of these eigenvectors can be computed in terms of $\omega$, $\xi$ and $D$ as,
\begin{align*}
    &\mathbf{q} = \left\langle\frac{\left(\frac{\sqrt{3}}{2}-\frac{i}{2}\right) \omega }{\sqrt{2} \sqrt{\omega ^2+1}},-\frac{\frac{1}{2}+\frac{i \sqrt{3}}{2}}{\sqrt{2} \sqrt{\omega
   ^2+1}},\frac{i \omega }{\sqrt{2} \sqrt{\omega ^2+1}},\frac{1}{\sqrt{2} \sqrt{\omega ^2+1}}\right\rangle\\
   &\mathbf{p} = Q\left\langle
   \frac{D \omega  \left(\frac{1}{2} \sqrt{3} \left(3 D \omega ^2+\xi
   \right)+\frac{3}{2} i \left(D \omega ^2-\xi \right)\right)}{3 D^2 \omega ^4+\xi
   ^2},\frac{1}{2} \left(1-i \sqrt{3}\right),\frac{D \omega  \left(\sqrt{3} \xi +3 i D
   \omega ^2\right)}{3 D^2 \omega ^4+\xi ^2},1
   \right\rangle,
\end{align*}
where 
\begin{equation*}
    Q = \frac{\sqrt{3 D^2 \omega ^4+\xi ^2}}{\sqrt{2} \sqrt{3 D^2 \omega ^2 \left(\omega
   ^2+1\right)+\xi ^2}}.
\end{equation*}

For simplicity of notation, write the reduced dimension version of \cref{eqn:TravellingWave} as,
\begin{equation*}
    \dot{\eta}_i = f_i(\bm{\eta}),
\end{equation*}
where $\bm{\eta} = \langle{u_1,v_1,u_2,v_2}\rangle$ and for $i=1,\dots,4$, $f_i$ is defined from \cref{eqn:TravellingWave}. Let $\bm{\eta}_0 = \left\langle{\tfrac{1}{3},0,\tfrac{1}{3},0}\right\rangle$ be the equilibrium point. Define $\mathbf{B}:\mathbb{R}^4 \times \mathbb{R}^4 \to \mathbb{R}^4$ componentwise as,
\begin{equation*}
B_i(\mathbf{r},\mathbf{s}) = \left.\sum_{j, k} \frac{\partial^2 f_i}{\partial \eta_j\partial \eta_k}\mathbf{r}_j\mathbf{s}_k\right\vert_{\bm{\eta} = \bm{\eta}_0}.
\end{equation*}
Define $\mathbf{C}:\mathbb{R}^4 \times \mathbb{R}^4\times\mathbb{R}^4 \to \mathbb{R}^4$ componentwise as, 
\begin{equation*}
C_i(\mathbf{r},\mathbf{s},\mathbf{w}) = \left.\sum_{j, k, l} \frac{\partial^3 f_i}{\partial \eta_j\partial \eta_k\partial \eta_l}\mathbf{r}_j\mathbf{s}_k\mathbf{w}_l\right\vert_{\bm{\eta} = \bm{\eta}_0}.
\end{equation*}
Lastly, define the complex inner product,
\begin{equation*}
    \langle{\mathbf{r},\mathbf{s}}\rangle = \sum_k \overline{\mathbf{r}}_k\mathbf{s}_k,
\end{equation*}
where $\overline{z}$ denotes the complex conjugate of the $z$. Then $\ell_1$ is computed as,
\begin{multline*}
    \ell_1 = \frac{1}{2\omega}\mathrm{Re}\left[
    \left\langle{\mathbf{p}, 
    \mathbf{C}(\mathbf{q},\mathbf{q},\overline{\mathbf{q}})}
    \right\rangle + \right.\\
    \left.\left\langle\mathbf{p},\mathbf{B}\left[\overline{\mathbf{q}},\left(2i\mathbf{I}\omega - \mathbf{J}_0\right)^{-1}\mathbf{B}(\mathbf{q},\mathbf{q}) \right]
    \right\rangle -
2\left\langle\mathbf{p},\mathbf{B}\left[\mathbf{q},\mathbf{J}_0^{-1}\mathbf{B}(\mathbf{q},\overline{\mathbf{q}})\right]\right\rangle
    \right].
\end{multline*}
Here, $\mathbf{I}$ is an identity matrix of appropriate size. Using Mathematica (see SI), it is straightforward to compute that,
\begin{align*}
&\left\langle{\mathbf{p}, 
    \mathbf{C}(\mathbf{q},\mathbf{q},\overline{\mathbf{q}})}
    \right\rangle = 0\\
&2\left\langle\mathbf{p},\mathbf{B}\left[\mathbf{q},\mathbf{J}_0^{-1}\mathbf{B}(\mathbf{q},\overline{\mathbf{q}})\right]\right\rangle = 0,
\end{align*}
leaving only the term,
\begin{equation*}
\left\langle\mathbf{p},\mathbf{B}\left[\overline{\mathbf{q}},\left(2i\mathbf{I}\omega - \mathbf{J}_0\right)^{-1}\mathbf{B}(\mathbf{q},\mathbf{q}) \right]
    \right\rangle,
\end{equation*}
to be evaluated. A human assisted computation with Mathematica (see SI) yields the expression,
\begin{equation*}
    \ell_1 = -\frac{3 \sqrt{3} \left(-3 D^2 \omega ^4+6 D \xi  \omega ^2+\xi ^2\right)}{4 \sqrt{\left(\omega ^2+1\right)^3 \left(3 D^2 \omega ^4+\xi ^2\right) \left(3 D^2 \omega ^2
   \left(\omega ^2+1\right)+\xi ^2\right)}}.
\end{equation*}

To prove this value is always negative, and thus that the limit cycle is always attracting, it suffices to show that,
\begin{equation*}
    -3 \sqrt{3} \left(-3 D^2 \omega ^4+6 D \xi  \omega ^2+\xi ^2\right) < 0,
\end{equation*}
for all allowable parameters. Computation is easier in terms of $\alpha$, $\beta$ and $D$ at this point. When we substitute in their definitions, we obtain the inequality,
\begin{equation*}
    \sqrt{3} \left(4 \alpha ^2-8 \alpha  \beta -4 \beta  (2 \beta +3)-3\right) \leq 0.
\end{equation*}
We can rewrite this as,
\begin{equation*}
\sqrt{3}\left[-3 + 4(\beta-\alpha)^2 - 12\beta(1+\beta)\right] \leq 0,
\end{equation*}
which implies,
\begin{equation*}
    (\beta - \alpha)^2 \leq \frac{12\beta(1+\beta)+3}{4}.
\end{equation*}
Solving the inequality for $\alpha$ yields the requirement that,
\begin{equation}
    \beta - \frac{\sqrt{3}}{2}\sqrt{1+4\beta+4\beta^2} \leq \alpha \leq \beta + \frac{\sqrt{3}}{2}\sqrt{1+4\beta+4\beta^2}.
    \label{eqn:LyapIneq}
\end{equation}
We know by our assumptions that $0 < \alpha < \beta$. Therefore, it suffices to show that the left-hand-side of the inequality is always less than zero. To prove this, note that for all $\beta > 0$ we have,
\begin{equation*}
   0 < \beta < \beta + \frac{\sqrt{3}}{2}\sqrt{1+4\beta+4\beta^2}.
\end{equation*}
Thus multiplying the left and right-hand sides of \cref{eqn:LyapIneq} yields,
\begin{equation*}
\left(\beta - \frac{\sqrt{3}}{2}\sqrt{1+4\beta+4\beta^2}\right)\left(\beta + \frac{\sqrt{3}}{2}\sqrt{1+4\beta+4\beta^2}\right) = -\left(2 \beta ^2+3 \beta +\frac{3}{4}\right) < 0.
\end{equation*}
Therefore, it follows that
\begin{equation*}
    \beta - \frac{\sqrt{3}}{2}\sqrt{1+4\beta+4\beta^2} < 0
\end{equation*}
and for all $0 < \alpha < \beta$, $\ell_1 < 0$ and thus the limit cycle is always attracting. 

\bibliography{HigherOrderSpatial}
\end{document}